\documentclass[twoside,leqno,twocolumn]{article}

\usepackage[letterpaper]{geometry}

\usepackage{ltexpprt}
\usepackage{hyperref}
\usepackage{amsmath,amssymb,bm,commath}

\usepackage{graphicx} 

\title{\Large Data mining the functional architecture of the brain's circuitry}
\author{Adam S. Charles\footnote{Department of Biomedical Engineering, Center for Imaging Science (CIS), Mathematical Institute for Data Science (MINDS), Kavli Neuroscience Discovery Institute (NDI), Data Science \& AI Institute, Johns Hopkins University, Baltimore, MD. Contact: adamsc@jhu.edu.}}
\date{}

\begin{document}

\maketitle


\begin{abstract} \small\baselineskip=9pt The brain is a highly complex organ consisting of a myriad of subsystems that flexibly interact and adapt over time and context to enable  perception, cognition, and behavior. Understanding the multi-scale nature of the brain, i.e., how circuit- and moleclular-level interactions build up the fundamental components of brain function, holds incredible potential for developing interventions for neurodegenerative and psychiatric diseases, as well as open new understanding into our very nature. Historically technological limitations have forced systems neuroscience to be \textit{local} in \textit{anatomy} (localized, small neural populations in single brain areas), in \textit{behavior} (studying single tasks), in \textit{time} (focusing on specific stages of learning or development), and in \textit{modality} (focusing on imaging single biological quantities). New developments in neural recording technology and behavioral monitoring now provide the data needed to break free of \textit{local} neuroscience to \textit{global} neuroscience: i.e., understanding how the brain's many subsystem interact, adapt, and change across the multitude of behaviors animals and humans must perform to thrive. Specifically, while we have much knowledge of the anatomical architecture of the brain (i.e., the hardware), we finally are approaching the data needed to find the functional architecture and discover the fundamental properties of the \textit{software} that runs on the hardware. We must take this opportunity to bridge between the vast amounts of data to discover this functional architecture which will face numerous challenges from low-level data alignment up to high level questions of interpretable mathematical models of behavior that can synthesize the myriad of datasets together.
\end{abstract}

\section{Introduction}

With the constant advancement of new neural recording technologies~\cite{jun2017fully,demas2021high,song2017volumetric,voleti2019real}, systems neuroscience has officially joined the era of big data~\cite{charles2020toward,benisty2022review}. Simultaneous recordings of tens of neurons has given way to hundreds and thousands~\cite{stevenson2011advances}, with millions of neurons no longer a pipe dream. Moreover, behavioral methods have significantly improved in parallel~\cite{wu2020deep,pereira2020sleap,nath2019using}, offering new avenues to train and monitor more complex behaviors, including freely moving animals during neural imaging~\cite{zong2022large,el2024chronic}, across organisms. With this explosion in data collection comes both opportunities to create a new data-driven view of neural function, but also challenges at every level from alignment to interpretability.

This opportunity will allow us to treat the brain as the interconnected system it is. For much of neuroscience history, studies at cellular resolution focused on local areas of the brain. Visual neuroscientists looked at the visual cortex, auditory processing was tested in auditory cortex, navigation in hippocampus, emotion and state in amygdala, etc. The brain, however, processes in parallel and distributed ways~\cite{rumelhart1986general}. Inactivating LIP---an area implicated in decision making---does not necessarily stop an animal from being able to make a decision~\cite{katz2016dissociated}. Studies in brain loss, and sensory loss redouble this observation, showing that the flexible brain substrate can move computations across the neural circuits to compensate for loss of tissue or to leverage unused resources~\cite{bedny2011language}. Brain-wide recordings can now provide unbiased cellular-level scans that let us map out the \textit{functional architecture}: where and how information spreads and transforms throughout the brain.

\begin{figure*}
        \includegraphics[width=\textwidth]{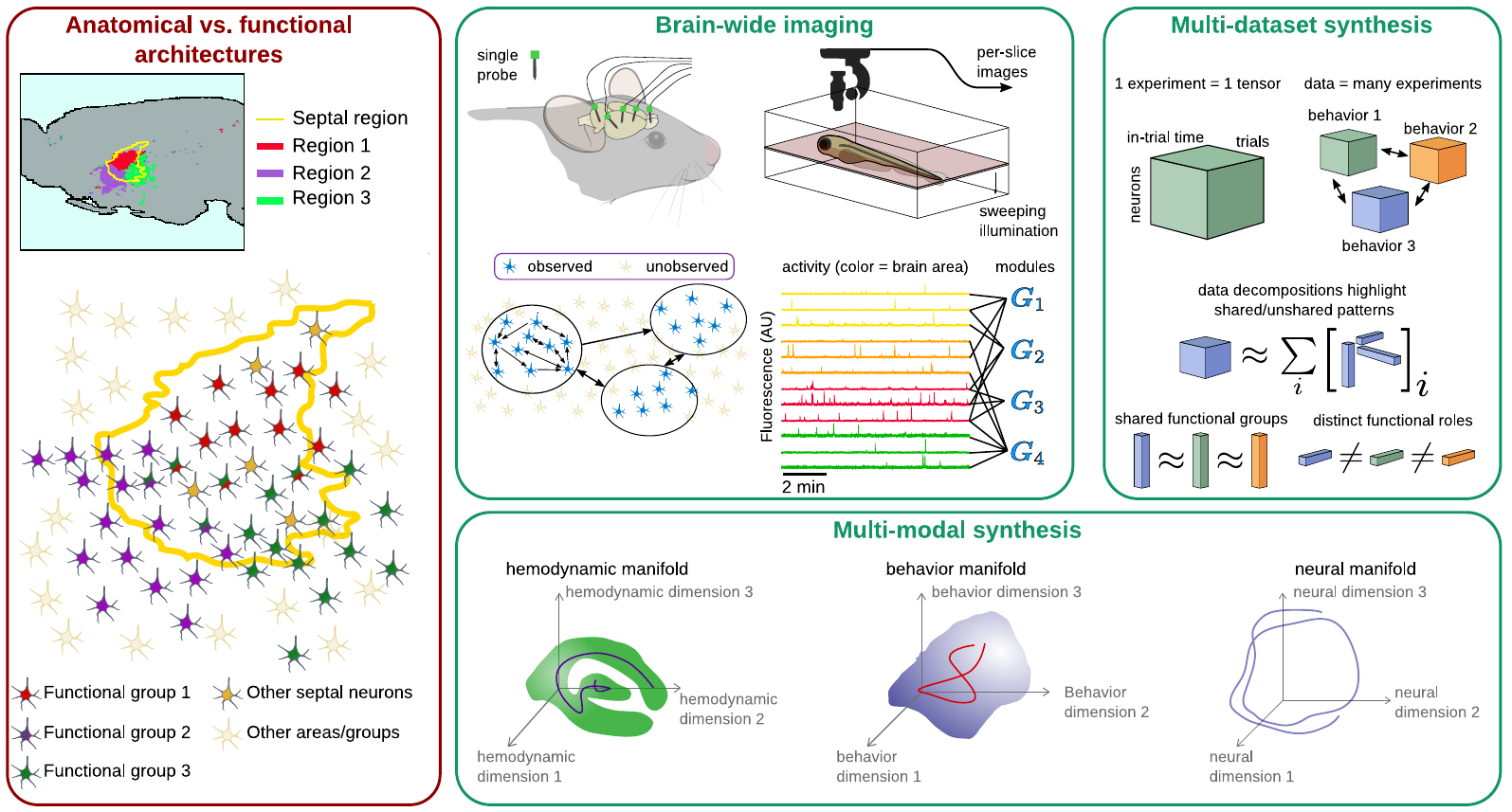}
        \caption{Challenges in discovering the brain's functional architecture. Left: many brain-wide recordings have found that functional relationships in the brain do not always follow the strict anatomical boundaries. Right: areas of advancing our ability to quantitatively find the functional architecture of the brain include synergizing across brain areas, across behavioral contexts, and across recording modalities.\vspace{-0.5cm}}
        \label{fig:overview}
\end{figure*}

A functional architecture would provide a roadmap to the general principles underlying the flexibility, robustness, and efficiency of neural computation. It will give us baselines for core functions that are necessary in healthy brains, which in turn will improve understanding of how observed activity changes in, e.g., neurodegenerative and psychiatric disorders. 
The current state-of-the-art is to identify brain regions---general anatomical areas---that have been linked to certain behavioral and cognitive aspects. However, as per the new global-brain observations, ``everything is everywhere''~\cite{steinmetz2019distributed,bondy2024coordinated,chen2024brain,koay2021sequential} and it is not clear if activity changes in a specific anatomical area must relate to a narrow set of functional deficits. 

To quantitatively map the functional architecture requires bridging a plethora of data: data taken across brain areas, across tasks, across modalities capturing different biophysical signals, and to combine these internal measurements with external observations, e.g., behavior. Bringing all these different datasets and synergizing across them all to produce a holistic view of a single computational system that adapts and learns is the primary challenge. Here I discuss a number of these challenges, specifically focusing on those challenges in higher-level mathematical modeling that will provide the language we need to merge all these datasets into a common framework.

\section{Rethinking systems-level models}
\begin{figure*}
        \includegraphics[width=\textwidth]{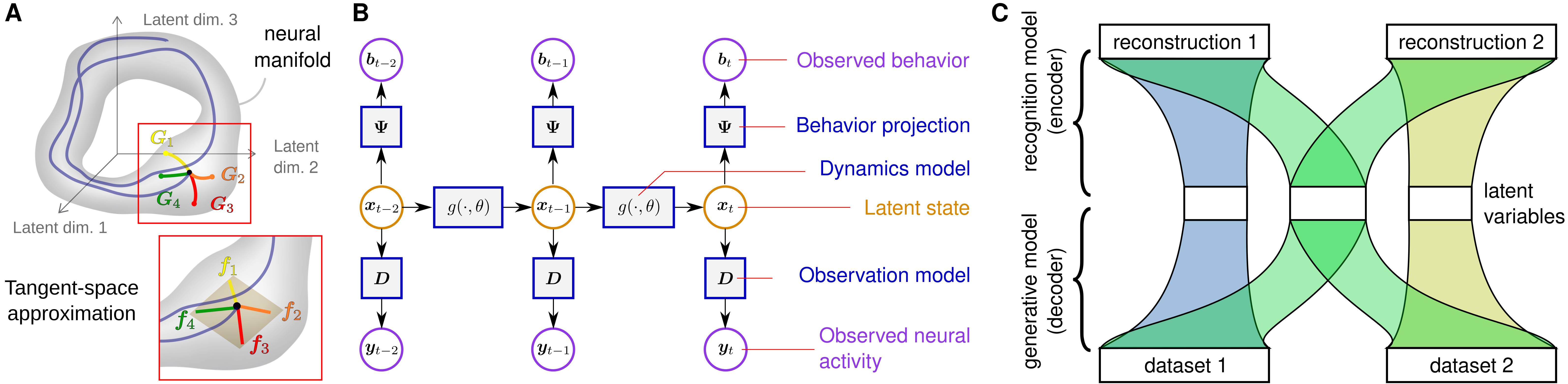}
        \caption{Example neural modeling approaches. A: defining the local geometry of the neural data can provide insight into the key features of the dynamics that define the flows across the manifold. B: dynamical latent states are a key  model for combining  brain and behavior into a single meaningful system-level model. C: relating disparate datasets requires finding shared and private sources of variance in an interpretable framework. \vspace{-0.5cm}}
        \label{fig:models}
\end{figure*}
When defining the brain's functional architecture, our models of neural dynamics must account for the multiple, distributed systems that comprise brain-wide computations. A primary challenge is that most of the fundamental models in neuroscience do not explicitly seek out these sub-systems. Instead the dominant mode of modeling dynamics is to consider every recorded unit as a single dimension in a shared state space over which the dynamics is not assumed to display preference. Put mathematically, if the time-series of unit $i$ is $x_i(t)$, then the vector $\bm{x}(t) = [x_1(t),\cdots,x_N(t)]^T$ is typically assumed to evolve as
$ \bm{x}_t = g(\bm{x}_{t-1}; \theta),$
i.e., a Markovian model where the parameters of the transition function $g(\cdot; \theta)$ are fit to data. Prominent examples of $g(\cdot; \theta)$ include linear dynamical systems~\cite{churchland2012neural}, switched linear dynamical systems~\cite{linderman2017bayesian}, and recurrent neural networks~\cite{yu2005extracting}. Non-Markovian versions of this core model also exist, including generalized linear models (GLMs)~\cite{pillow2008spatio} and RNNs with long-term connections, e.g., Long-Short Term Memory (LSTM) and General Recurrent Unit (GRU) networks. In all these models, the full system is treated as co-evolving at the same time-scale. Implicitly, the learned network connectivity can be used to identify disjoint systems, e.g., when the linear transition matrix of an RNN is block-diagonal~\cite{perich2020rethinking}, however given that even simple RNNs can have a plethora of local minima, there is a danger of over-interpreting the parameters that are fit to finite, noisy data.  

The ability to explicitly learn modular dynamics is key to identifying the functional architecture's constitute components. Mining the data in an interpretable way is thus critical, and the modularity must be built directly into our models. A number of recent efforts have begun to learn these modular dynamics. One example is the decomposed linear dynamical systems (dLDS)~\cite{mudrik2024decomposed}. This approach treats dynamics as paths on a manifold $\bm{x}(t) \in \bm{M}\subset\mathbb{R}^N$. The approach then considers the tangent spaces at each point on the manifold $\mathcal{T}(x(t))$, and finds a set of linear operators $\{\bm{G}_k\}_{k=1,\ldots,K}$ that, when applied to each point, span their respective tangent spaces, i.e., $\mathcal{T}(x(t)) \subseteq \mbox{span}(\bm{G}_1\bm{x}(t),\cdots,\bm{G}_K\bm{x}(t))$ for all $\bm{x}(t)$. Each operator describes a distinct set of interactions that are driven by how the neural state transverses the manifold. To ensure that these interactions are meaningful, dLDS further assumes that the use of the operators is \textit{sparse}, i.e., the cardinal directions of each tangent space's basis align with the paths taken by the neural state.  
Similar approach also take the manifold partitioning approach~\cite{tafazoli2024building,acosta2023relating}. The general idea of finding compositional descriptions of geodesic trajectories in the manifold interpretation of neural data appears to be a powerful tool to identify modular structure in the brain.


\section{Challenges in multi-dataset analysis}

Synthesizing data across the many contexts under which the brain has been recorded is another key challenge in finding the brain's functional architecture. Specifically, data-driven discovery is driven by patterns and correlations in the data, and if a specific task recruits two systems simultaneously at all times, the ensuing analysis will never be able to differentiate between those systems. For example, in a reaction task, the visual and motor stimuli activate together in a single burst of activity. Meanwhile, in another task, the two systems might be used quite differently, e.g., in a decision making task where the visual system is separated in time and context from the motor reporting. 

In fact, the existence of distinct systems is likely driven by the need to flexibly perform multiple tasks. Recent work has studied machine learning systems, specifically recurrent neural networks, to understand the internal dynamics that emerge when a single systems is trained to perform multiple tasks~\cite{driscoll2024flexible,vafidis2024disentangling}. These studies have found that in fact multi-task RNNs seem to develop internal modules that are only visible under extensive additional analysis (e.g., fixed-point analyses).   

In neuroscience, finding such modules thus relies on the synthesis of datasets across tasks, i.e., across experimental sessions, individual animals, and even labs. Thus alignment in terms of neuron-to-neuron alignment is impossible, and instead systems-level alignment must be performed. For example, recent efforts has aimed at leveraging graph-based approaches~\cite{charles2022graft} for multi-matrix (or multi-tensor) decompositions where shared factors can be constrained to represent similar functions, providing a key link between datasets~\cite{mudrik2023sibblings,mudrik2024creimbo}. This link can be, e.g., in neurons if highly overlapping population are observed across contexts/tasks. Alternatively, the link can be in the trial structure for similar behaviors across animals. While promising, there are core assumptions about the shared trial or neural structure in these approaches, and different assumptions will likely be needed in each analysis.    

\section{Challenges in interpretability}

Thus far I have avoided mention of modern artificial neural network (ANN) based AI and ML that now form the centerpieces of large-scale data mining in other applications.   
This is because the black box nature of ANNs prevents the types of interpretability necessary for scientific discovery.
Scientists need to find relationships between variables that extrapolate our understanding and recommend broader relationships between objects of study. Thus analysis methods such as manifold analysis, causal analysis, and simpler linear systems tend to still be used extensively despite the increased ease of training and deploying ANNs. 
More specificly, ANNs sacrifice interpretability for expressivity. Thus while they \emph{interpolate} well on the domain of the data provided, they do not \emph{extrapolate} beyond those confines. 

This is not to say that no ANN tools can aid scientists in learning about the brain. 
For example, in variational autoencoders, sparsity and independence in the latent layers has been used to promote a disentangling of the data representation that produces more interpretable representations~\cite{burgess2018understanding,geadah2024sparse}. Moreover, emerging approaches in explainable AI, e.g., Shapley analysis~\cite{teneggi2022fast}, and others~\cite{bharti2024sufficient,dwivedi2023explainable}, can identify key features in the dataset that drive the ANNs. Often, however, these latter approaches do not provide the extrapolation necessary for scientific discovery. Instead they are self-contained explanations of the original data.   

In dynamical systems, there are similar trade-offs in  expressivity and interpretability. For example, large RNNs with LSTM and GRU nodes can predict future data to very high accuracy, however the learned RNN parameters are not unique and thus limited in the insights that can be gleaned about the system's internal interactions. Moreover, nonlinear systems are now being described in latent spaces, i.e., low-dimensional representations of the neural data $\bm{x}(t) = \bm{Phi}(\bm{z}(t))$. When the neural data is a  non-linear function of the latents, i.e., $\bm{\Phi}(\cdot)$ is nonlinear. Also including nonlinear dynamics results in an unidentifiability up to an invertible function $h(\cdot)$. Specifically, if $\bm{z}(t) = g(\bm{z}(t-1))$ and $\bm{x}(t) = \bm{\Phi}(\bm{z}(t))$ then we can define $\widetilde{\bm{\Phi}}= \bm{\Phi}\circ h$, $\widetilde{\bm{z}(t)} = h^{-1}(\bm{z}(t))$ and $\widetilde{g} = h^{-1}\circ g \circ h$, yielding a second solution $\widetilde{\bm{\Phi}}$, $\widetilde{g}$, $\widetilde{\bm{z}}(t)$ that describe the data equally well. Thus, theoretical advances are also needed to understand when such combinations of extremely expressive models creates statistically unidentifiable models.   

\section{Challenges in multi-modal data}

Finally, the brain is not just a set of pyramidal neurons exchanging electrical spikes. There is a rich biophysical infrastructure of neurons, astrocytes, vasculature, and a plethora of neurotransmitters that modulate neural function beyond direct connections~\cite{randi2023neural,yezerets2024decomposed}. These additional structures  are a part of the brain's computation and should be included in the functional architecture. 
Each of these additional signals can be measured using different technologies and resolutions: fMRI or functional ultrasound for hemodynamics, optical imaging for neurotransmitters, etc. Moreover, behavioral monitoring provides yet another mode that captures the environment and eventual output of the brain: the actions taken in the world.  

Sythesizing across data types require a single model that describes multiple modalities. This model class, called data fusion, joint modeling, or multi-model modeling (depending on the community), is often defined by a latent state $\bm{z}$ such where each mode $\bm{x}_1$ and $\bm{x}_2$ can be ``read out'' from the latent state, e.g., $\bm{x}_1 = f_1(\bm{z})$ and $\bm{x}_2 = f_2(\bm{x}_2)$. Approaches to extract the shared latent state span from linear Canonical Correlations Analysis (CCA)~\cite{hardoon2004canonical} to deep-learning based extensions (deepCCA)~\cite{andrew2013deep} and non-parametric manifold learning approaches~\cite{lederman2018learning}. While such methods can identify one such joint representation, the relationship of multiple datasets is often more nuanced with some information in one mode that is not in the other. E.g., voltage data has temporal precision not present in hemodynamics, while hemodynamics covers a larger spatial extent. Thus, a more complete view requires separating the latent space into shared and private information, i.e., $\bm{z} \rightarrow \{\bm{z}_s, \bm{z}_1,\bm{z}_2\}$, where $\bm{z}_s$ are the variables representative of the shared information while $\bm{z}_1$ and $\bm{z}_2$ are the private information representations for $\bm{x}_1$ and $\bm{x}_2$, respectively. 

ANN architectures, e.g., cross-encoders with private paths, can learn private latents. However, their high  expressibility often causes private information to leak into the shared variables and vice versa. This leakage, while not necessarily damaging in engineering applications, can cause erroneous scientific conclusions about shared brain function. Recent methods leverage ``butterfly'' architectures that pair multiple cross-encoders with adversarial predictions to minimize such leakage~\cite{koukuntla2024unsupervised}. 

Beyond static comparisons, another active area that can start shedding light into the functional architecture is the joint encoding of dynamical time-series information~\cite{sani2021modeling,gondur2023multi}. The general conceptual framework of the above holds, however these methods are often phrased as generative models where the latent representation evolves dynamically in time, and both brain and behavior data can be extracted from the same states. The same challenges apply to the dynamical systems framework in terms of making sure the shared and private information are fully disentangled. 

\section{Conclusion}

I aim here to lay out a key opportunity in mining the depths of now-available neural data: discovering the brain's functional architecture. In this endeavor, the field will have to so solve at a minimum the mentioned challenges, specifically 1) the synthesis of data collected across brain areas, behaviors, and modalities, 2) the synthesis of brain and behavior data, and 3) the development of interpretable AI that goes beyond the explainable AI currently used in engineering applications. 

In the emerging solutions, one emerging theme is the importance of data geometry, specifically going beyond topology and into how the curvature and tangent spaces relate to dynamics. Another theme is the importance of statistically independent representations, which is related to the sparsity that is enjoying a rebound in use from its ability to induce interpretability into regression-type problems. These advances and more will hopefully soon provide new insights into brain function.   

\bibliographystyle{plain}
\bibliography{refs}

\end{document}